\documentclass[conference]{IEEEtran}
\IEEEoverridecommandlockouts
\usepackage{cite}
\usepackage{amsmath,amssymb,amsfonts}
\usepackage[noend]{algpseudocode}
\usepackage{algorithm}
\usepackage{graphicx}
\usepackage{textcomp}
\usepackage{xcolor}
\usepackage{placeins}
\usepackage{comment}
\def\BibTeX{{\rm B\kern-.05em{\sc i\kern-.025em b}\kern-.08em
    T\kern-.1667em\lower.7ex\hbox{E}\kern-.125emX}}
\title{A Bayesian Hierarchical Model for Generating Synthetic Unbalanced Power Distribution Grids
\thanks{This study was financed, in part, by the São Paulo Research Foundation (FAPESP), Brasil. Process Number \#2023/07634-7. This study was financed, in part, by the São Paulo Research Foundation (FAPESP), Brasil. Process Number \#2021/12220-1.}
}

\author{
\IEEEauthorblockN{Henrique O. Caetano\\}
\IEEEauthorblockA{Department of Electrical and Computing Engineering \\
University of São Paulo (EESC/USP) - São Carlos, Brazil\\
henriquecaetano1@usp.br}
\and
\IEEEauthorblockN{Rahul K. Gupta}
\IEEEauthorblockA{School of Electrical Engineering \& Computer Science\\
Washington State University, USA \\ 
rahul.k.gupta@wsu.edu}
\and 
\IEEEauthorblockN{Marco Aiello}
\IEEEauthorblockA{Department of Service Computing\\ 
University of Stuttgart - Stuttgart, Germany\\
marco.aiello@iaas.uni-stuttgart.de}
\and
\IEEEauthorblockN{Carlos Dias Maciel}
\IEEEauthorblockA{Faculty of Engineering and Science\\
São Paulo State University (UNESP) - Guaratinguetá, Brazil\\
carlos.maciel@unesp.br }
}

\usepackage{etoolbox}
\makeatletter
\patchcmd{\@maketitle}
  {\addvspace{0.5\baselineskip}\egroup}
  {\addvspace{-2.0\baselineskip}\egroup}
  {}
  {}
\makeatother

\begin{document}

\setlength{\textfloatsep}{1pt}
\setlength{\floatsep}{1pt}
\setlength{\intextsep}{1pt}

\maketitle

\begin{abstract}
The real-world data of power networks is often inaccessible due to privacy and security concerns, highlighting the need for tools to generate realistic synthetic network data. Existing methods leverage geographic tools like OpenStreetMap with heuristic rules to model system topology and typically focus on single-phase, balanced systems, limiting their applicability to real-world distribution systems, which are usually unbalanced. This work proposes a Bayesian Hierarchical Model (BHM) to generate unbalanced three-phase distribution systems learning from existing networks. The scheme takes as input the base topology and aggregated demand per node and outputs a three-phase unbalanced system. The proposed scheme achieves a Mean Absolute Percentage Error (MAPE) of less than $8\%$ across all phases, with computation times of 20.4 seconds for model training and 3.1 seconds per sample generation. The tool is applied to learn from publicly available SMART-DS dataset and applied to generate European 906 and IEEE-123 systems. 
We demonstrate the transfer learning capability of the proposed tool by leveraging a model trained on an observed system to generate a synthetic network for an unobserved system. Specifically, the tool is trained using the publicly available SMART-DS dataset and subsequently applied to generate synthetic networks for the European 906-bus system and the IEEE 123-bus system.
This tool allows researchers to simulate realistic unbalanced three-phase power data with high accuracy and speed, enhancing planning and operational analysis for modern power grids.
\end{abstract}

\begin{IEEEkeywords}
Power distribution systems, synthetic test cases, bayesian inference, three-phase unbalanced systems, 
\end{IEEEkeywords}

\section{Introduction}
The transition towards sustainable energy systems are introducing significant changes to the power distribution systems (DS) and has brought attention to their robust operation, planning, and analysis tools for  \cite{li2022review}. These tools often require 
% several changes to power distribution systems (DS). Increasing penetration of renewable energy sources, distributed generation, and the growing complexity of electricity demand highlight the need for robust planning, operation, and analysis tools for DS \cite{li2022review}. Consequently, there has been an increasing attention for robust planning, operation and analysis of such systems, and 
access to data from real systems - such as the topology and network parameters as well as the evolution of its demand and generation power for the proper modeling and methodology validation. However, due to data confidentiality and security concerns, the data set for real-world power systems may not be accessible \cite{Caetano2024}, making it difficult for researchers to test and develop new methods and algorithms for modern DS planning and control. In this context, synthetic power distribution systems have emerged as a promising solution, providing realistic and versatile data sets that mimic the characteristics of DS in the real world without disclosing sensitive information \cite{Caetano2024, Ali2023, Li2020, mateo2020building}.

%paragraph 2: saying the contradiction between openstreetmaps data and the lack of real power data for customers
Existing works on synthetic DS often use open platforms like OpenStreetMaps (OSM) \cite{OpenStreetMap} and relies on heuristic methods to estimate the network topology, parameters and demand profile \cite{Cakmak2022, Ali2023, Li2020, Gupta2021}. %. These tools offer publicly accessible data about the physical layout of distribution systems, including poles, substations, and feeder routes \cite{Cakmak2022, Ali2023, Li2020, Gupta2021}. 
% However, these georeferenced-based methodologies face limitations when it comes to estimating synthetic power data, particularly regarding the daily demand of users, such as active and reactive power. 
Typically, the approach involves associating residential buildings from OSM with a proportional relationship between the active power of each node and the size of nearby residences or buildings. Although this simplifies the process of collecting electrical properties, it raises concerns when evaluating critical aspects such as power quality, reliability under failure scenarios, and system performance. 
% This discrepancy underscores a key challenge: while georeferenced data provides essential information for system topology, the absence of real power data for customers restricts the creation of comprehensive models that are vital for effective analysis, planning, and ensuring the operational integrity of power distribution systems.

%paragraph 3: saying that no work so far in the literature have been dealing with 3-phase synthetic power systems
% In order to overcome the power estimation simplification that exits in georefernced-based tools, another set of works in the literature have been using 
Another approach is based on the statistical tools that characterize the distribution of power demand and component location for synthetic DS. For example, the work done in \cite{Schweitzer2017} considered DS as a graph, where buses are nodes and lines are edges, each with various properties, and leverages this structure to search for emerging statistical patterns. The work in \cite{Wang2022} presented a graph-theoretic approach to create a synthetic cyber-power network based on user inputs and system domain constraints. The work carried out by \cite{Caetano2024} uses a Bayesian model to allocate network components such as normally open and normally closed switches. 

%paragraph 4: gaps to be filled
An important gap in the current literature is the lack of methodologies capable of generating data for unbalanced 3-phase systems. In real-world power distribution systems, especially during peak demand periods, imbalances between phases are common occurrence \cite{kersting2018distribution}. Accurate modeling of these imbalances is crucial, as they impact the performance of the system, including power quality, stability of the system, and the effectiveness of control strategies. Furthermore, real DS often exhibit hybrid configurations, where some nodes are single-phase and others are 3-phase. The ability to model these hybrid systems, with varying phase configurations, requires sophisticated statistical models that can account for both the unbalance in 3-phase sections and the mixed nature of the system. Therefore, developing methodologies capable of generating realistic data for unbalanced and hybrid systems is essential to improve the accuracy of synthetic power system models and ensure that they are applicable to a wide range of real-world scenarios. Existing methods such as \cite{mateo2020building} have devised a scheme to produce unbalanced distribution grids, but they use a rule-based approach, which may not reflect real-world networks.

This work aims to address the gaps mentioned above by developing a probabilistic tool learning from the existing unbalanced networks, which is then used in the Bayesian Hierarchical Model (BHM) to sample the distribution of phase type and per phase demand for a given topology of distribution system.
% Bayesian framework to generate unbalanced three-phase distribution systems that learn from existing networks. The scheme takes as input the base topology and aggregate demand per node and outputs a three-phase unbalanced system. 
The key contributions are
\begin{enumerate}
    % \item To the best of our knowledge, no existing methodology has successfully created 3-phase synthetic DS, an essential aspect for accurately simulating real-world power grids. 
    \item We propose a probabilistic tool for quantifying key characteristics of the 3-phase unbalanced systems i.e., the degree of unbalance in the phases or the proportion of 3-phase nodes as well as the demand distribution per phase with respect to the distance from the feeder.
    % which are crucial for realistic system analysis and performance evaluation;
    \item We demonstrate transfer learning capability of the trained model, i.e., the ability to learn the probabilities of one system and transfer this knowledge to others with different scales, allowing us to utilize the data on existing real or synthetic networks.
    % This approach not only enables the adaptation of the model to systems of varying sizes but also allows for the incorporation of expert knowledge to improve accuracy or simulate hypothetical scenarios. 
    \item We demonstrate fast sampling capability by leveraging BHM, enhancing computational efficiency while maintaining robust probabilistic representations. 
\end{enumerate}

% \begin{comment}
The paper is structured as follows: Section \ref{sec:framework} introduces the proposed framework for generating synthetic unbalanced three-phase DS. Section \ref{sec:results} presents the key results, emphasizing model validation and performance evaluation. Lastly, Section \ref{sec:conclusions} concludes the work, highlighting its contributions, limitations, and potential future directions.
% \end{comment}

\section{Synthetic Network Generation Framework}
\label{sec:framework}
\begin{figure}[!ht]
\label{fig:flowchart}
\centering
\includegraphics[width=\linewidth]{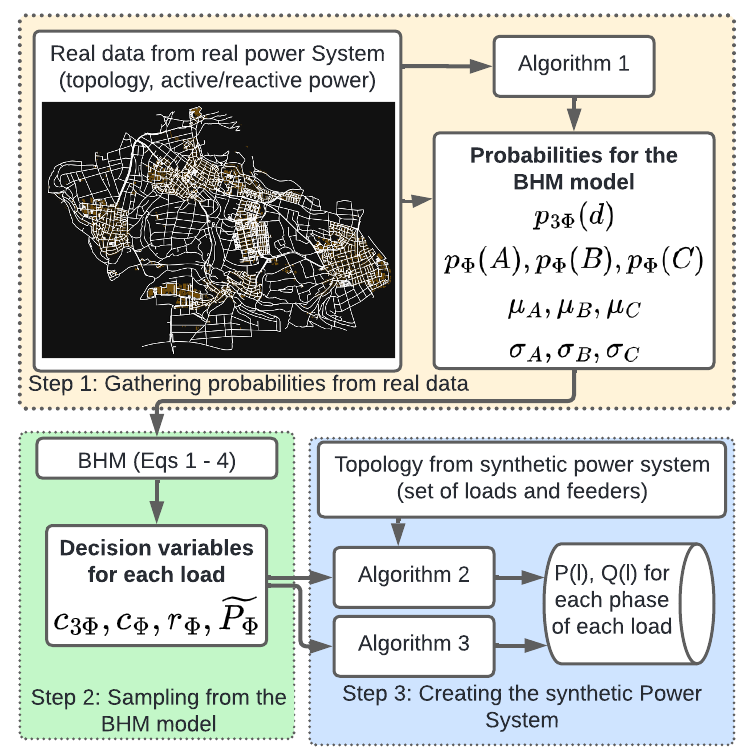}
\caption{Flowchart of the proposed methodology.}
\label{fig:sensitivity_analysis}
\end{figure}

We propose a Bayesian Hierarchical Model (BHM) for generating the synthetic network which takes as an input the probabilities of various grid parameters, i.e., probability of existing a single or three-phase node, distribution of the demand per phase, and its distribution with the distance from the substation. Once the BHM is obtained, we use that to sample out synthetic networks. The scheme is summarized in the flow chart shown in Fig. \ref{fig:flowchart}.
% illustrates the proposed methodology through a flowchart. 
%The process begins with Step 1 (Sec.~\ref{subsec:step_1}), where real-world network dataused to determine probabilities related to power demand and phase balance. In Step 2 (Sec.~\ref{subsec:step_2}), these probabilities serve as inputs to construct a Bayesian Hierarchical Model (BHM), which define the decision variables for the synthetic distribution network. Finally, Step 3 (Sec.~\ref{subsec:step_3}) utilizes these decision variables along with the topology of the synthetic power system to generate the final three-phase demand (active and reactive power) for all loads in the synthetic system. Each step is described in the following.

\subsection{Estimating Probabilities from the Real data}
\label{subsec:step_1}
One of the primary goals of the proposed approach is to leverage the existing real-world data to sample out realistic unbalanced distribution systems. Therefore, 
% generate realistic power profiles and phase allocation for a synthetic power system, even when the topology of the synthetic system differs from the original. To achieve this goal, 
% algorithm 
% \ref{alg:normalized_3_phase_prob} is utilized to define $p_{3\Phi(d)}$, employing a normalized distance metric where all values are scaled between 0 and 1. This normalization ensures that the same probability distribution can be consistently applied to synthetic power systems, even when they differ in scale. 
% Other variables are derived directly from the data, with the mean and standard deviation of active power being calculated under the assumption that demand follows a normal distribution—a premise that has been supported in previous studies \cite{Caetano2024}. Additionally, the methodology is flexible enough to accommodate alternative distributions for active power, as discussed in detail in Subsection \ref{subsec:step_2}.
we estimate a set of probabilities related to the unbalancedness of the distribution system using this real data, they are defined as follows:
\begin{enumerate}
\item $p_{3\Phi}(d)$: The probability of a load having three phases is modeled as a function of its normalized distance \( d \) from the feeder. This probability is computed using Algorithm~\ref{alg:normalized_3_phase_prob}, which employs a normalized distance metric. 
\item $p_{\Phi}(A), p_{\Phi}(B), p_{\Phi}(C)$: Probabilities that a single-phase load is allocated to phase A, phase B, and phase C, respectively.
\item $\mu_{\Phi}$: Mean value of the active power (in kW), representing the total aggregated demand across all phases for each individual load;
\item $\sigma_{\Phi}$: Standard deviation of the active power, representing the total aggregated demand across all phases for each individual load
\item $\overline{r_{\Phi}(A)}, \overline{r_{\Phi}(B)}, \overline{r_{\Phi}(C)}$: Percentage of active power that is allocated for each phase, compared to the total load demand. For example: if a load has 25kW for phase A, 15kW for phase B and 10kw for phase C, then $\overline{r_{\Phi}(A)}=0.5, \overline{r_{\Phi}(B)}=0.3$ and $\overline{r_{\Phi}(C)}=0.2$
\end{enumerate}

% Consider that the real data input consists of the following: The topology of the power system - loads and feeders, as well as their connections; the active and reactive power of each phase of each load (in the case of single phase loads, 2 out of 3 values will be set to 0). 

% \begin{algorithm}[!ht]
% \caption{Define Normalized 3-phase probability as function of distance from feeder}\label{alg:normalized_3_phase_prob}
% \hspace*{\algorithmicindent} \textbf{Input:} Set of loads ($\mathcal{L}$) and Feeders ($\mathcal{F}$) of the real power system
% \begin{algorithmic}[1]
% \State $p_{3\Phi}(d)=0$, $\forall d \in \{0,1\}$ 
% \For{$l \in \mathcal{L}$}
% \State $f$ = \texttt{FeederFromLoad}($l$) \Comment{Feeder connected to load}
% \State $\Bar{d}$ =\texttt{Distance}($l$,$f$)/\texttt{MaxDistance}($f$) \Comment{Normalized distance}
% \If{LoadIs3Phase($l$)=True}
% \State $p_{3\Phi}(d)=p_{3\Phi}(\Bar{d})+1$ 
% \EndIf
% \EndFor
% \State $p_{3\Phi}(d)=p_{3\Phi}(d)/length(L)$, $\forall d \in \{0,1\}$ \Comment{All values sum to 1}
% \end{algorithmic}
% \end{algorithm}

\begin{algorithm}[!ht]
\caption{Normalized 3-phase probability as function of distance from feeder}\label{alg:normalized_3_phase_prob}
\hspace*{\algorithmicindent} \textbf{Input:} Set of loads ($\mathcal{L}$) and feeders ($\mathcal{F}$) from real network
\begin{algorithmic}[1]
\State $p_{3\Phi}(d)=0$, $\forall d \in \{0,1\}$ 
\For{$l \in \mathcal{L}$}
\State Find the feeder $f$ connected to the load $l$
\State Find Normalized distance between $l$ and $f$ as $\Bar{d}$  \hspace{2em}  = (Distance between $l$ and $f$) / (Length of the Feeder $f$)
\If{Load $l$ is 3-Phase }
\State $p_{3\Phi}(d)=p_{3\Phi}(\Bar{d})+1$ 
\EndIf
\EndFor
\State $p_{3\Phi}(d)=p_{3\Phi}(d)/\texttt{length}(\mathcal{L})$, $\forall d \in \{0,1\}$ \Comment{All values sum to 1}
\end{algorithmic}
\end{algorithm}
%%%%%%%%%%
\subsection{Bayesian Hierarchical Model}
\label{subsec:step_2}
%In this step, the probabilities extracted in the previous section are utilized to build a BHM. The random variables modeled within the BHM are specifically designed to capture the characteristics of an unbalanced 3-phase power system. These include determining whether a load is single-phase or three-phase, identifying the specific phase to which a single-phase load is assigned, and distributing the power demand across all phases for three-phase loads. The final output of this step consists of the active power and reactive power vectors ($P(l)$) and $Q(l)$) for $l \in \mathcal{L}$, each containing three values for phases A, B and C, respectively.
% Both will be a list of three values, where the first value represents the power for phase A, the second for phase B and the third for phase C.
The load type, single-phase or three-phase, is modeled using a Bernoulli distribution, as
\begin{align}
    & c_{3\Phi} \sim \texttt{Bernoulli}(p_{3\Phi}(d)) \label{eq:num_of_phases}
\end{align}
\noindent where, $c_{3\Phi}$ takes the value of 0 and 1 for single- and three-phase loads, respectively. This choice is appropriate because the Bernoulli distribution is well-suited for representing binary outcomes of a single experiment, where the question being addressed is whether the load is single-phase (yes) or three-phase (no). 
% as shown in Eq. (\ref{eq:num_of_phases}). 

For single-phase loads, the choice of phase is modeled using a Categorical distribution, as
\begin{align}
    c_{\Phi} \sim \texttt{Categorical}(p_{\Phi}(A), p_{\Phi}(B), p_{\Phi}(C)) \label{eq:phase_choice}
\end{align}
\noindent where, $c_{\Phi}$ takes the value 0, 1, or 2 if phase A, B, or C is selected, respectively. This distribution is chosen because it is a discrete probability distribution that models the outcomes of a random variable with \( k = 3 \) categories. The probability of each category is explicitly defined as \( p_{\Phi}(A/B/C) \) for phase A/B/C.

For three-phase loads, the phase ratio is modeled using a Dirichlet distribution, as 
\begin{align}
    & r_{\Phi}(A), r_{\Phi}(B), r_{\Phi}(C) \sim \texttt{Dir}(\overline{r_{\Phi}(A)}, \overline{r_{\Phi}(B)}, \overline{r_{\Phi}(C)}) \label{eq:power_ratio} 
\end{align}
% shown in Eq. (\ref{eq:power_ratio}). 
Here, $r_{\Phi}(A/B/C)$ represents the proportion of power demand allocated to phase A/B/C. The symbol $r_{\Phi}(A/B/C)$ indicates that $(r_{\Phi}(A/B/C))\%$ of the load's power demand is assigned to phase A/B/C. The Dirichlet distribution is chosen for this variable as its output is restricted between 0 and 1, while ensuring their sum equals 1. This property makes it suitable for modeling the phase type proportions, as it naturally aligns with the concept of probabilities in a $k$-way categorical event, in this specific case $k=3$ for three phases.

Finally, the total demand of a given load is modeled as a Truncated Normal (to allow only positive values) as 
% shown in Eq. (\ref{eq:total_demand}). 
\begin{align}
    % & c_{3\Phi} \sim \texttt{Bernoulli}(p_{3\Phi}(d)) \label{eq:num_of_phases} \\ 
    % & c_{\Phi} \sim \texttt{Categorical}(p_{\Phi}(A), p_{\Phi}(B), p_{\Phi}(C)) \label{eq:phase_choice} \\ 
    % & r_{\Phi}(A), r_{\Phi}(B), r_{\Phi}(C) \sim \texttt{Dir}(\overline{r_{\Phi}(A)}, \overline{r_{\Phi}(B)}, \overline{r_{\Phi}(C)}) \label{eq:power_ratio} \\ 
    & \widetilde{P_{\Phi}} \sim \texttt{TruncatedNormal}(\mu_{\Phi},\sigma_{\Phi}), \, \forall \Phi \in \{A,B,C\} \label{eq:total_demand}
\end{align} 
\noindent where each phase A/B/C has its own mean ($\mu$) and variance ($\sigma$). Normal distribution is chosen based on previous work found in the literature \cite{Caetano2024}. However, it is important to emphasize that the BHM is sampling-based, which allows it to be generalized to accommodate any distribution for the active power of loads without requiring modifications to the upper layers - Eqs. (\ref{eq:num_of_phases})-(\ref{eq:power_ratio})—of the model.

\subsection{Creating the Synthetic Power System}
\label{subsec:step_3}
After defining the decision variables in Eqs. (\ref{eq:num_of_phases})–(\ref{eq:total_demand}), the next step involves utilizing these variables to sample the demands per node for each phase, i.e., $P(l)$ and $Q(l)$. This process is outlined in Algorithm \ref{alg:defining_load_type_and_power}. If a load is determined to be three-phase (line 2), the power for each phase is allocated based on the calculated ratios (line 3). Conversely, if the load is single-phase (line 4), the power is assigned entirely to one of the phases based on the phase choice variable (lines 5 - 9), while the remaining phases have their demands set to zero. Finally, the reactive power is calculated on line 11, where the power factor for the whole network ($PF$) is modeled as a simple three possibilities as shown in (\ref{eq:power_factor}), where $u$ is a random uniform distribution in the closed interval $[0,1]$, as previously used in the literature \cite{Cakmak2022}.
\begin{algorithm}[!ht]
\caption{Define number of phases, active and reactive powers based on probability distributions}\label{alg:defining_load_type_and_power}
\hspace*{\algorithmicindent} \textbf{Input:} Set of loads ($L$) of the synthetic power system; $c_{3\Phi}$, $c_{\Phi}$, $r_{\Phi}$, $\widetilde{P_{\Phi}}$
\begin{algorithmic}[1]
\For{$l \in L$}
\If{$c_{3\Phi}(l) = 1$} \Comment{Load is 3 phase}
\State $P(l)=(r_{\Phi}(A) \cdot \widetilde{P_{\Phi}},r_{\Phi}(B) \cdot \widetilde{P_{\Phi}},r_{\Phi}(C) \cdot \widetilde{P_{\Phi}})$
\Else 
\If{$c_{\Phi}=0$}
\State $P(l)=(\widetilde{P_{\Phi}},0,0)$
\EndIf
\If{$c_{\Phi}=1$}
\State $P(l)=(0,\widetilde{P_{\Phi}},0)$
\EndIf
\If{$c_{\Phi}=2$}
\State $P(l)=(0,0,\widetilde{P_{\Phi}})$
\EndIf
\EndIf
\EndFor
\State $Q_{l}(i)=P_{l}(i) \cdot \texttt{tan}(\texttt{arccos}(PF)), \forall i \in \{0,1,2\}$
\end{algorithmic}
\end{algorithm}

\begin{equation}
\label{eq:power_factor}
\
    PF = 
\begin{cases}
    0.85,& \text{if } 0 < u \leq 0.1649\\
    0.90,& \text{if } 0.1649 < u \leq 0.27\\
    0.95,& \text{otherwise} \\
\end{cases}
\    
\end{equation}

\begin{comment}
This entire process of defining decision variables, coupled with the conditional logic implemented in Algorithm \ref{alg:defining_load_type_and_power}, was developed within the BHM framework. By leveraging the built-in functions of the PyMC library \cite{pymc_ref_2016}, from the `pm.math` package, it was possible to integrate complex conditional structures directly into the BHM. This approach combines the flexibility of handling intricate rules, such as the if-else conditions, with the computational efficiency and sampling advantages inherent to the BHM framework. This ensures accurate modeling while maintaining scalability for large datasets and intricate scenarios.
\end{comment}

\subsection{Filtering samples for phase consistency}
\label{subsec:sampling_filtering}
One of the steps required for proper phase allocation in an unbalanced distribution system is phase consistency, where if a phase is allocated to a bus, it must also be allocated for every other bus on the path between it and the feeder. This is also extended for buses with multiple phases (i.e. AB or BC), where both phases must be allocated on the other buses in the path. Different approaches are used in the literature for phase consistency, for example, \cite{postigo2020phase} developed a heuristic method, \cite{gupta_molzahn-phasebal_milp} developed a mixed-integer-linear-program (MILP) to ensure phase consistency. 

Here, we utilize a rule-based approach, described in Algorithm \ref{alg:phase_consistency}. Phases(n) represent the phases of node $n$ (i.e.$\{A\}$, $\{A,B\}$ or $\{A,B,C\}$. ShortestPath( G,$n_{1}$,$n_{2}$ returns the edges that represent the path between nodes $n_{1}$ and $n_{2}$, where each edge is represented as a pair of nodes. NumPhases($n$) returns the number of phases (1, 2 or 3) assigned to node $n$. The algorithm ensures phase consistency in a power distribution network by iteratively checking and adjusting the phases of nodes along the paths from leaf nodes to the feeder, ensuring that downstream phases consistent with the upstream phases.

\begin{algorithm}[!ht]
\caption{Achieving phase consistency accross the network}\label{alg:phase_consistency}
\hspace*{\algorithmicindent} \textbf{Input:} Graph G(V,E) representing the network, phases assigned for each node: Phases($n$) $\forall n \in V$
\begin{algorithmic}[1]
\For{$n_{l}$ in LeafNodes(G)}
\State path=ShortestPath(G,$n_{f}$,$n_{l}$)
\For{($n_{d}$,$n_{u}) \in path$}
\If{Phases($n_{d}$) $\nsubseteq$ Phases($n_{u}$)}
\If{NumPhases($n_{u}$) $\geq$ NumPhases($n_{d}$)}
\State Phases($n_{u}$) = Phases($n_{u}$) $\cup$ Phases($n_{d}$)
\Else
\For{$n_{2}$ in ShortestPath(G,$n_{f}$,$n_{u}$})
\State Phases($n_{2}$) = Phases($n_{u}$) $\cup$ Phases($n_{d}$)
\EndFor
\EndIf
\EndIf 
\EndFor
\EndFor
\end{algorithmic}
\end{algorithm}

% create a flowchart better addressing each step in the procedure. highlight how algorithm 1 takes real data and outputs a probability distribution, and how algorithm 2 takes such data and applies in a totally different system (where we know only the topology), allowing for better transfer learning

\section{Results and Discussion}
\label{sec:results}
%to-do:
% 1) figure 2 should be smaller
% 2) show typical values of other probabilities (other than Fig 2). probably show on a table
% 5) for the feasibility I should show the percentage of feasible solutions, also show how much it take to create so many samples (and show that is fast). also mention proper filtering within the sampling phase is good for next steps
% adapt all figures, description etc to account for the new variable names and symbols I proposed in the framework section
We use python Pymc4 library \cite{pymc_ref_2016} to model the BHM. The PandaPower package \cite{pandapower_ref_2018} was used to simulate the synthetic power distribution system. The Networkx library \cite{networkx_ref_2008} was used to calculate graph properties.

\subsection{Fitting Real-world Data to the BHM}
\label{subsec:fitting_data}
The first step is to fit the proposed BHM using publicly available data. We use SMART-DS datasets \cite{oedi_2981}, which is publicly available large-scale US electrical distribution synthetic system. We specifically used data from the Austin region in its base scenario and data from all substations were considered. 

\begin{comment}
We extract the following data on
\begin{itemize}
    \item Active (kW) and reactive (kvar) power for each phase (A, B and C) of each node;
    \item Geographical location of each load point and feeder for computing the distance between each load and its connected feeder.
\end{itemize}
These data are used to estimate the probabilities of a node having three phases with distance from the feeder, and distribution of the active power allocation per phase for a three-phase node. 
\end{comment}

\begin{figure}[!ht]
\centering
\includegraphics[width=\linewidth]{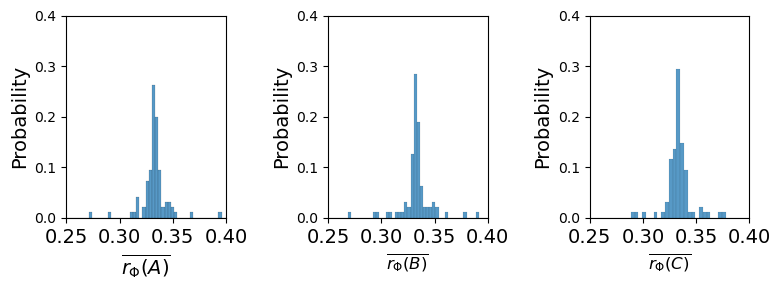}
\caption{Histogram showing the distribution of the percentage of active power allocated to each phase for 3-phase loads, derived from collected data.}
\label{fig:phase_ratio}
\end{figure}

Figure \ref{fig:phase_ratio} presents the histogram that illustrates the proportion of active power allocated to each phase for all loads. The distribution is centered around 0.33 (or 1/3), which is expected since the majority of 3-phase loads in this particular dataset are balanced.

\begin{figure}[!ht]
\centering
\includegraphics[width=\linewidth]{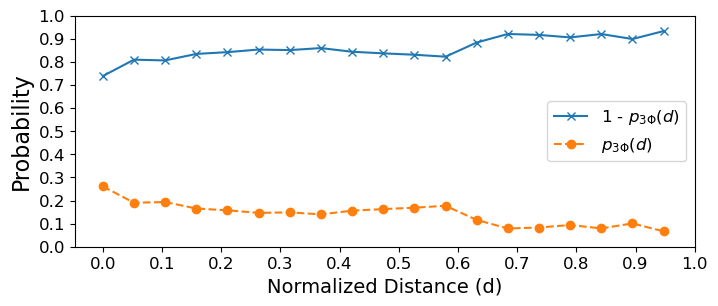}
\caption{Probability of a load being 3-phase ($p_{3\Phi(d)}$) as a function of the normalized distance $d$, derived from the SMART-DS dataset.}
\label{fig:prob_3_phase_loads}
\end{figure}

Figure \ref{fig:prob_3_phase_loads} illustrates the variation of $p_{3\Phi}(d)$ as a function of the normalized distance from the feeder. The results clearly show a decrease in the likelihood of three-phase loads as the distance increases, a trend consistent with real-world systems, where three-phase loads are typically located closer to the feeder due to topological considerations. This spatial variation in the probability of three-phase loads underscores the value of the proposed approach, which not only probabilistically evaluates these changes but also leverages real data to accurately fit these probabilities.

Finally, the posterior mean for the parameters related to phase choice of single phase loads are as follows: $p_{\Phi}(A)=0.3350$; $p_{\Phi}(B)=0.3296$; $p_{\Phi}(C)=0.3354$ 

%%%%%%%%%%%%%%%%
\begin{comment}
\subsection{Sensitivity Analysis}
One of the primary objectives of the proposed methodology was to establish a probabilistic approach capable of generating unbalanced three-phase synthetic power systems. This section conducts a sensitivity analysis on the main categorical random variables related to the three-phase balance of the generated samples: $r_{\Phi}(A), r_{\Phi}(B)$ and $r_{\Phi}(C)$. Figure \ref{fig:sensitivity_analysis} illustrates how variations in these probabilities affect the overall active power across each phase.

The analysis reveals a clear relationship between the probability values and the overall active power for each phase. When the probabilities are balanced at (1/3, 1/3, 1/3), the active power across all phases is equal, approximately 0.4 kW. This outcome occurs because three-phase loads are assigned equal active power. Conversely, in the non-uniform scenario (e.g., (1, 0, 0) for Phase A, (0, 1, 0) for Phase B, and (0, 0, 1) for Phase C), all active power is allocated to a single phase.

\begin{figure}[!ht]
\centering
\includegraphics[width=\linewidth]{figures/phases_proportion_sensitivity_analysis.png}
\caption{Impact of the probabilities of the Phases Proportion for each phase ($r_{\Phi}(A), r_{\Phi}(B), r_{\Phi}(C)$) on the mean active power for each phase, averaged across all samples generated by the BHM model}
\label{fig:sensitivity_analysis}
\end{figure}  
\end{comment}

\subsection{Creating synthetic power systems from user input}

In this set of results, the main objective is to evaluate how the proposed BHM model can generate synthetic power data based on the combination of: 1) the initial topology of a given power system and 2) the input of the user, in the form parameters for the probability distributions described in Equations (\ref{eq:phase_choice}) - (\ref{eq:total_demand}). Two distinct scenarios are modelled:
\begin{itemize}
    \item Balanced scenario: $\overline{r_{\Phi}(A)}=\overline{r_{\Phi}(B)}=\overline{r_{\Phi}(C)}=1/3$
    \item Unbalanced scenario: $\overline{r_{\Phi}(A)}=0.1$; $\overline{r_{\Phi}(B)}=0.6$; $\overline{r_{\Phi}(C)}=0.3$
\end{itemize}

For each scenario, a total of 1000 samples were generated using the proposed BHM.

\begin{figure}[!ht]
\centering
\includegraphics[width=\linewidth]{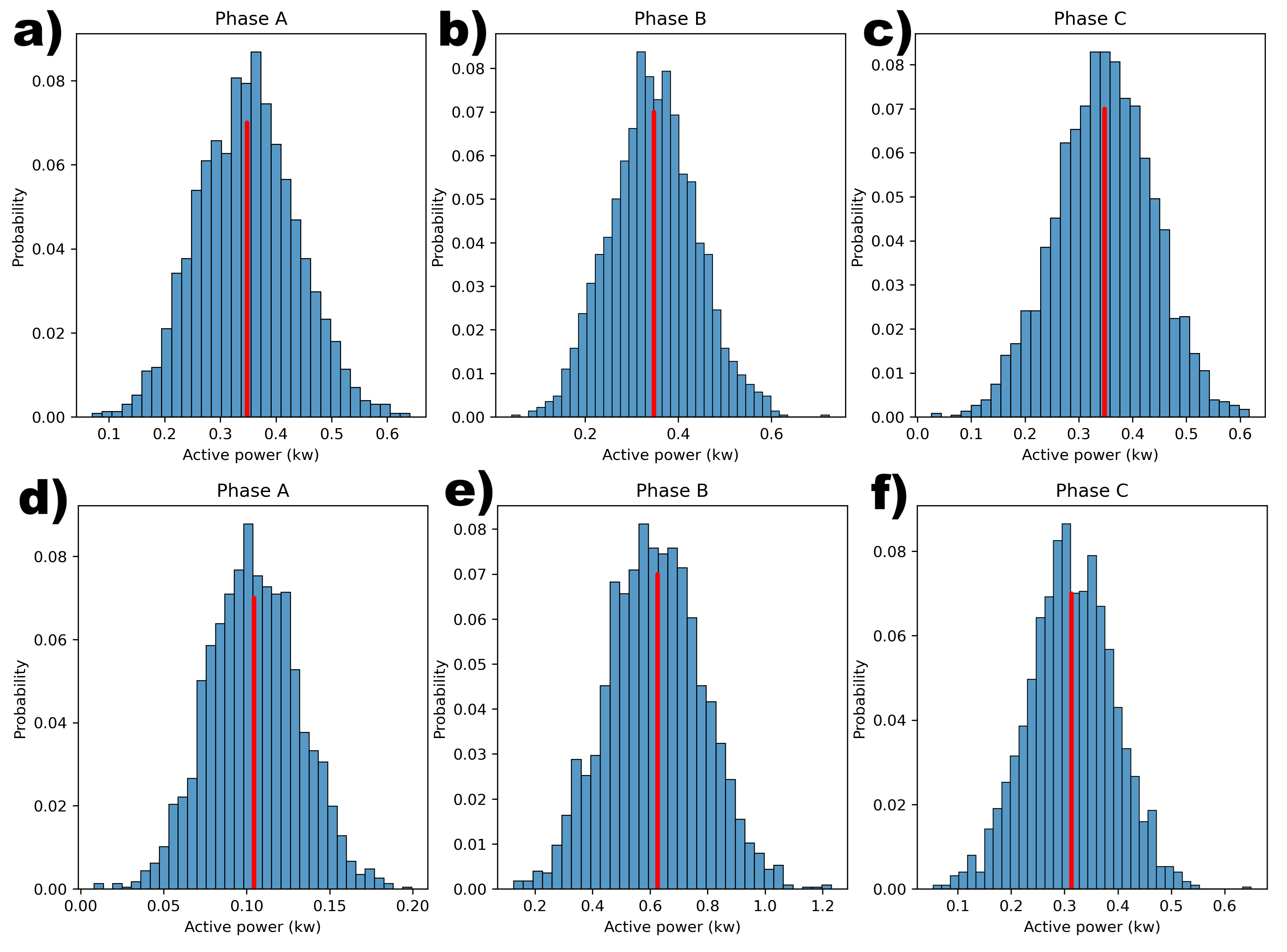}
\caption{Comparison between user input (red line) and BHM model-generated samples (blue bars) for active power distribution across all loads and three network phases, presented for both balanced (upper) and unbalanced (lower) scenarios.}
\label{fig:comparing_samples_and_user_input}
\end{figure}

Figure \ref{fig:comparing_samples_and_user_input} presents a comparison between user input values and the samples generated by the BHM model for active and reactive power of single-phase loads in all phases. The results indicate that the mean values remain consistent across the three phases reflecting a (un)balanced network, depending on the simulated scenario. In terms of the BHM-generated samples, both the mean and the most frequent values (i.e., those with the highest probabilities) closely align with the user input mean, demonstrating the model's capability to capture the expected user-defined behavior in both balanced and unbalanced scenarios. Some generated samples display deviations from the mean, highlighting the robustness of the proposed model. By modeling active and reactive power as random variables, the BHM enables the creation of scenarios that not only align with the user-defined mean but also capture the inherent uncertainties present in real-world systems.

\subsection{Using transfer learning to create synthetic power systems}
\label{subsec:transfer_learning}
An important advantage of the proposed methodology is its ability to perform transfer learning—leveraging data from one system (referred to as System A) to generate synthetic data for another system (referred to as System B). However, it is crucial to assess how well the generated synthetic data aligns with the real data from System B. In most real-world applications, the real data for System B is typically unavailable, which is the primary motivation for generating synthetic data. Nevertheless, in this specific scenario, the real data for System B is known, allowing for an evaluation of the methodology's performance through the following steps:
\begin{enumerate}
    \item Data from system A is used to feed the BHM, we use SMART-DS dataset \cite{oedi_2981} as system A.
    \item The BHM is used to sample synthetic power data from system B. In this case, the European Low Voltage Network \cite{ieeeResourcesx2013} topology was chosen as the starting one. This system was originally a generic 0.416 kV network serviced by one 0.8 MVA MV/LV transformer station. The network supplies 906 LV buses and 55 1-PH loads. The network layout is mostly radial
    \item The synthetic power data is compared with the real power data from system B.
\end{enumerate}

It is important to emphasize that, in this process, the real data from System B is not used to train or inform the BHM. Instead, it serves solely as a benchmark to assess the accuracy and reliability of the generated synthetic data in approximating the true system behavior.

\begin{figure}[!ht]
\centering
\includegraphics[width=0.9\linewidth]{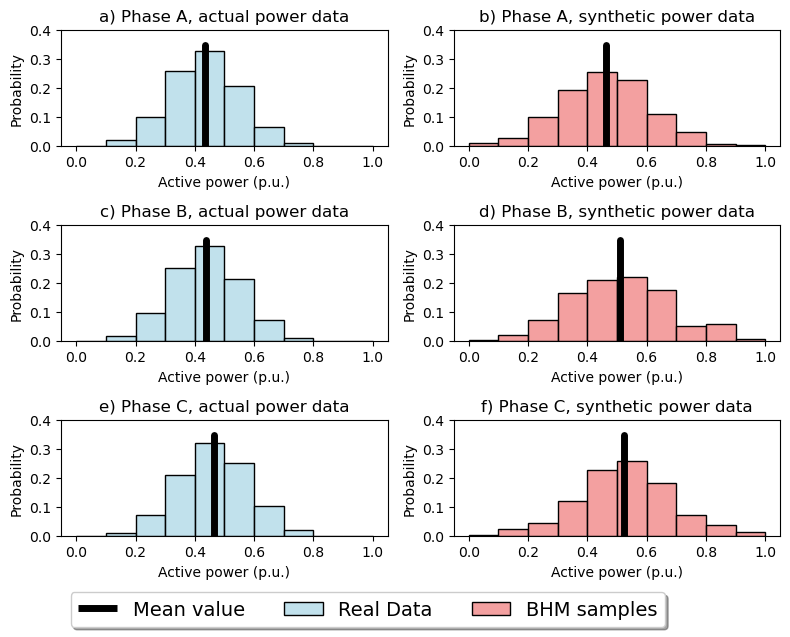}
\caption{Assessment of the proposed approach's transfer learning capability. The histograms depict the active power distribution for the same power system.}
\label{fig:transfer_learning_active_power_histogram}
\end{figure}

Figure \ref{fig:transfer_learning_active_power_histogram} presents the histograms for the active power distribution across all three phases for all loads in the system. The histograms compare the real power data (left) and the synthetic power data (right), along with their respective mean values. The figure demonstrates that the distributions of both datasets are highly similar across all three phases, visually validating the ability of the proposed methodology to replicate the statistical properties of the original system. 

Table \ref{table:mape_error_transfer_learning} presents the Mean Average Percentage Error (MAPE) between the actual and synthetic datasets. The table confirms the accuracy of the synthetic data, showing how closely the means of the datasets align. The MAPE remains below $8\%$ for the active power of all three phases, emphasizing the reliability of the proposed approach in capturing the key characteristics of the original system.

Furthermore, the parameters associated with the probability of a load being three-phase and the phase assignment for single-phase loads are compared between the real power system and the generated one. A summary of the results is presented in Table \ref{table:three_phase_assignment_transfer_learning}, demonstrating that the proposed approach generates synthetic data with phase assignment characteristics closely matching those of the real system. Notably, the approach accurately estimates that approximately 11\% of the loads are three-phase.

% \vspace{1em}

\begin{table}[!ht]
\caption{Mean value of the active power for each phase, presented for both the real power data and the synthetic power data, along with the final MAPE of the synthetic data estimation.}
\label{table:mape_error_transfer_learning}
\begin{center}
\resizebox{\columnwidth}{!}{
\begin{tabular}{|c|c|c|c|}
\hline
\textbf{Phase} & \textbf{Mean (kW) - real data}&\textbf{Mean (kW)- synthetic data}&\textbf{MAPE} \\
\hline
A & 0.454 & 0.423 & $7.292\%$ \\
B & 0.434 & 0.455 & $4.715\%$ \\
C & 0.464 & 0.483 & $3.916\%$ \\
\hline
\end{tabular}}
\end{center}
\end{table}

\begin{table}[!ht]
\centering
\caption{Comparison of probabilities for three-phase connection ($p_{3\Phi}$) and phase assignments for phases A, B, and C ($p_{\phi}(A/B/C)$) derived from both real power data and synthetic power data.}
\label{table:three_phase_assignment_transfer_learning}
\begin{center}
\resizebox{0.6\columnwidth}{!}{
\begin{tabular}{|c|c|c|}
\hline
\textbf{Parameter} & \textbf{Real Data}&\textbf{Synthetic Data} \\
\hline
$p_{3\Phi}$ & 0.1111 & 0.1242 \\
$p_{\Phi}(A)$ & 0.3350 & 0.3819 \\
$p_{\Phi}(B)$ & 0.3296 & 0.3454 \\
$p_{\Phi}(C)$ & 0.3354 & 0.2728 \\
\hline
\end{tabular}}
\end{center}
\end{table}

\begin{figure}[!ht]
\centering
\includegraphics[width=\linewidth]{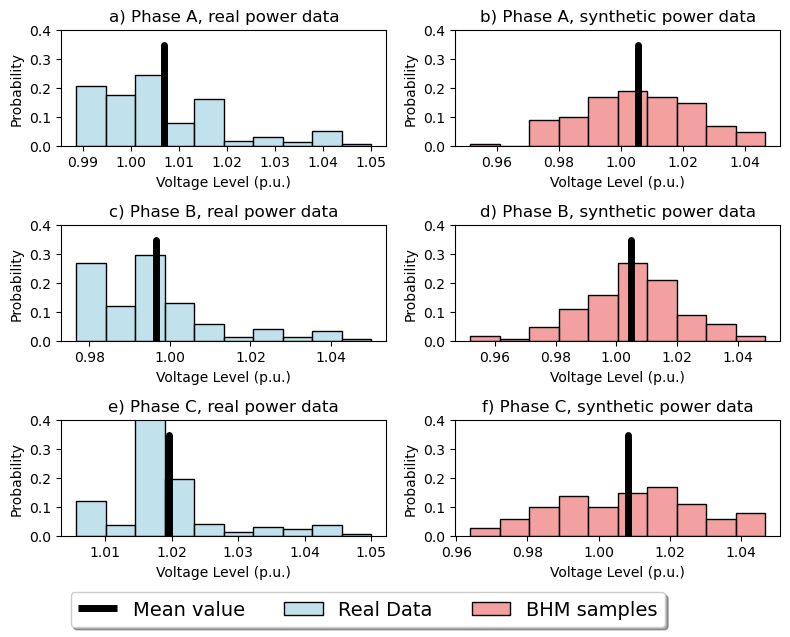}
\caption{Assessment of the proposed approach's transfer learning capability. The histograms depict the voltage level (p.u.) distribution for the same power system.}
\label{fig:transfer_learning_voltage_level_histogram}
\end{figure}

Power quality is evaluated by comparing the voltage levels (in p.u.) of the synthetic data with those of the real system. The histograms for each phase are displayed in Fig. \ref{fig:transfer_learning_voltage_level_histogram}. Standard regulations state that voltage levels should be around 0.9 to 1.1 p.u. for all phases \cite{Gupta2021}. As it can be seen by Figure \ref{fig:transfer_learning_voltage_level_histogram}, none of the samples have a voltage level lower than 0.95 nor higher than 1.04, and thus the voltage levels are respected.

\subsection{Validation of Phase Consistency Algorithm}
\label{subsec:ieee_123_test_case}

\begin{figure}[!ht]
\centering
\includegraphics[width=0.9\linewidth]{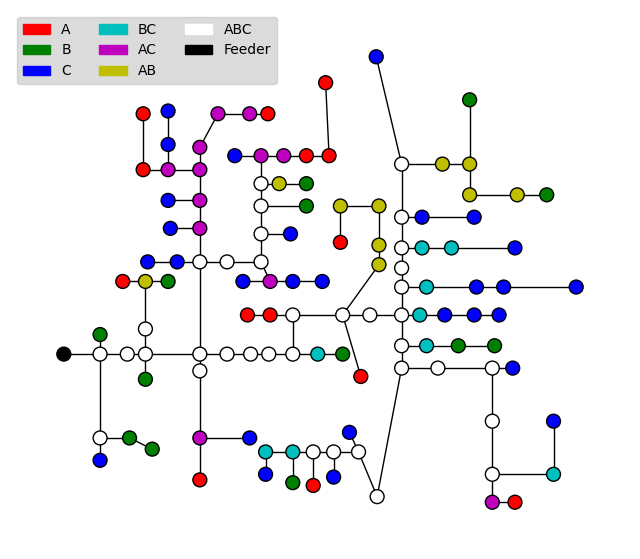}
\caption{Phase allocation results on the IEEE 123-bus test case \cite{iee_123_ref}, illustrating the assigned phases for each bus (A, B, C, or their combinations).}
\label{fig:123_test_case}
\end{figure}

We validate the network sampling and phase consistency for the phase allocation on a different system B, which is highly unbalanced: the IEEE 123-bus benchmark system \cite{iee_123_ref}. Note that transfer learning is still applied here, with the only difference being that system B is now the IEEE 123 system instead of the European low-voltage system. 
For this purpose, the BHM trained in Section \ref{subsec:fitting_data} is utilized on the IEEE 123-bus system, then we apply the phase consistency algorithm described in Subsection \ref{subsec:sampling_filtering}. The resulting phase allocations for each bus are illustrated in Figure \ref{fig:123_test_case}. The results indicate that phase consistency is maintained across the entire network, with a well-distributed balance among the allocated phases (A, B, C, and combinations thereof).

\section{Conclusions}
\label{sec:conclusions}
% The increasing complexity of DS, driven by modern technologies and growing demand, coupled with data restrictions due to security concerns, underscores the need for tools capable of generating synthetic, ready-to-use data for power system analyses. 
% This work addresses a key gap in the literature, where no existing tool has successfully 
This work proposes a probabilistic tool for generating synthetic networks for unbalanced three-phase power distribution systems. We developed a Bayesian hierarchical model consisting of a framework of statistical layers and rules to accurately generate and allocate single-phase and three-phase unbalanced load active and reactive power data. These outputs are based on key probability distributions that can be fitted using existing publicly available real-world data. We demonstrated the transfer capability of the trained algorithm, where we showed that a model learned from SMART-DS dataset can be used to sample a different unbalanced network, such as European 906 and IEEE 123 system.
The proposed methodology demonstrated the ability to generate samples with low error—achieving a MAPE below $8\%$—and low computational time, making it suitable for real-time operational tools.

Despite these achievements, the methodology has limitations that suggest directions for future research. Key next steps include refining the allocation of physical components such as generators, storage units, and switches; implementing a filtering mechanism during the sampling phase to ensure the generation of only feasible outputs, embedding the phase consistency algorithm within the training and sampling process, and estimating active and reactive power for active elements such as generators and prosumers supplying power.

% mention proper filtering within the sampling phase is good for next steps

\bibliographystyle{IEEEtran}
\bibliography{IEEEabrv,ref}

\end{document}